\documentstyle[12pt]{article}
\begin{document}
\title{Optimal Eavesdropping in Quantum Cryptography. II. Quantum Circuit}

\author{Robert B. Griffiths\thanks{Electronic mail: rgrif+@cmu.edu}\ \
and Chi-Sheng Niu\thanks{Electronic mail: cn28+@andrew.cmu.edu}\\ Department of
Physics\\ Carnegie Mellon University\\ Pittsburgh, PA 15213, U.S.A.}

\date{}
\maketitle

\begin{abstract}
	It is shown that the optimum strategy of the eavesdropper, as described
in the preceding paper, can be expressed in terms of a quantum circuit in a way
which makes it obvious why certain parameters take on particular values, and
why obtaining information in one basis gives rise to noise in the conjugate
basis.
\end{abstract}


\section{Introduction}
\label{s1}

	The preceding paper \cite{fel97} discusses the maximum information
which an eavesdropper, Eve, can obtain for a given error rate between Alice,
who sends signals, and Bob, their legitimate recipient, in the context of the
BB84 cryptographic scheme \cite{bb84}.  In the present paper we show that Eve's
optimum strategy, discussed in Sec.~III of the preceding paper, can be embodied
in a simple quantum circuit, of the type proposed for quantum
computation~\cite{qc}, together with appropriate initial states of the two
qubits which constitute Eve's probe, and suitable final measurements.

	The quantum circuit is extremely simple: it involves only two gates, of
the controlled-not variety, although adding a third gate could be advantageous
under some circumstances (discussed towards the end of Sec.~\ref{s3}). When it
is analyzed using the same consistent history methods we used earlier
\cite{gn96} to simplify the final Fourier transform in Shor's factorization
algorithm \cite{shor}, it is immediately obvious how the parameters in Eve's
initial state are related to the error rates, both errors in the transmission
from Alice to Bob, and the errors which determine the mutual information
between Alice and Eve.

	The results of the preceding paper which are essential for
understanding the present one are summarized in Sec.~\ref{s2} below. The
quantum circuit is described in analyzed in Sec.~\ref{s3}, and a brief summary
is presented in Sec.~\ref{s4}.

\section{Errors and Information}
\label{s2}

	In the BB84 scheme, Alice transmits a signal using a qubit described by
a two-dimensional Hilbert space; for example, the polarization of a photon, or
the spin of a spin half particle.  The kets $|x\rangle $ and $|y\rangle $ form
an orthonormal basis of this space, and
\begin{equation}
 |u\rangle = (|x\rangle + |y\rangle )/\sqrt{2}, \quad |v\rangle = (|x\rangle -
|y\rangle )/\sqrt{2},
\label{euv}
\end{equation}
form an alternative (conjugate) basis.  Alice chooses one of these bases at
random, and then transmits one of the basis vectors, also chosen at random.
Her qubit, hereafter denoted by $a$, is intercepted by Eve and made to interact
with a probe consisting of two qubits, which we shall call $e$ and $f$.  After
this, Eve sends $a$ on to Bob, who measures it in one of the two bases, again
chosen at random. Eventually Alice announces publicly the basis she used for
transmission of the signal, and in those cases in which Bob measured in the
same basis (the other cases are of no interest, for Alice and Bob discard the
results), Eve, who now knows the basis Alice employed, measures the qubits in
her probe in order to estimate which signal Alice sent.

	In Sec.~II of the preceding paper it was shown that the average
information (in the Shannon sense) $I_{xy}$ which Eve obtains about Alice's
signal when the latter uses the $xy$ basis is bounded by
\begin{equation}
I_{xy}\leq\mbox{$1\over2$}\,\phi\Bigl[ 2\sqrt{D_{uv}\,(1-D_{uv})}\,\Bigr],
\label{Ixy}
\end{equation}
where
\begin{equation} 
\phi(z)=(1+z)\ln(1+z)+(1-z)\ln(1-z),
\label{ephi}
\end{equation}
and $D_{uv}$ is the error rate produced by the interaction with Eve's probe
when Alice transmits and Bob measures in the $uv$ basis.  Similarly, when Alice
sends a signal in the $uv$ basis, the average information $I_{uv}$ which Eve
can gain is bounded by a similar inequality
\begin{equation}
I_{uv}\leq\mbox{$1\over2$}\,\phi\Bigl[ 2\sqrt{D_{xy}\,(1-D_{xy})}\,\Bigr],
\label{Iuv}
\end{equation}
with $D_{xy}$ the error rate from Alice to Bob when they employ the $xy$ basis.

\section{Quantum Circuit}
\label{s3}

	The quantum circuit used in Eve's optimum strategy is shown in
Fig.~\ref{fgx}.  The three horizontal lines represent three qubits thought of
as moving from left to right as time increases.  The top line is Alice's qubit
$a$ on its way to Bob, while the two lower lines show qubits $e$ and $f$
representing Eve's probe.  These interact with Alice's qubit through two
controlled-not gates, labeled 1 and 2 and indicated by solid vertical lines.
The dashed vertical lines indicate where the qubits are at two specific times
which we will want to refer to later.

	The two diagrams in Fig.~\ref{fgx} represent the same quantum circuit,
but show its action in two different bases: the $xy$ basis in (a) and the $uv$
basis in (b).  By $xy$ basis we now mean the basis of the full
eight-dimensional Hilbert space corresponding to qubits $a$, $e$, and $f$, with
each qubit in either an $|x\rangle $ or a $|y\rangle $ state.  Thus the basis
vectors are of the form $|aef\rangle =|xxy\rangle ,|xyx\rangle ,$ and so forth.
Similarly, the $uv$ basis is constituted by vectors of the form $|aef\rangle
=|vuv\rangle $, and so forth; each of the qubits is in either the state
$|u\rangle $ or the state $|v\rangle $, where these are related to $|x\rangle $
and $|y\rangle $ through (\ref{euv}).

	In Fig.~\ref{fgx}(a), the $xy$ basis, gate 1 is a unitary
transformation which when applied to qubits $a$ and $e$, with the left letter
in each ket referring to $a$ and the right to $e$, yields
\begin{equation}
  |xx\rangle \to |xx\rangle , \quad |xy\rangle \to |xy\rangle , \quad
|yx\rangle \to |yy\rangle , \quad |yy\rangle \to |yx\rangle .
\label{gate1a}
\end{equation}
That is, if $a$ is in state $|y\rangle $, $e$ is flipped from $|x\rangle $ to
$|y\rangle $ or vice versa, whereas if $a$ is in state $|x\rangle $, $e$
remains unchanged; in either case, $a$ retains its original value.  Such a gate
is called ``controlled-not'', because flipping a bit corresponds to logical
negation, and whether or not the {\it target} qubit $e$ is flipped depends on
the state of the {\it control} qubit $a$.  Note that $f$ is not involved in
gate 1, so (\ref{gate1a}) can be extended to the eight basis vectors of the
full Hilbert space by inserting a third letter in each of the kets to represent
the state of $f$, the same letter on each side of the arrow: $|yxx\rangle \to
|yyx\rangle , |yxy\rangle \to|yyy\rangle ,$ and so forth.

	Gate 2 in Fig.~\ref{fgx}(a) is another controlled-not operation, but
now $f$ is the control qubit and $a$ the target qubit, whereas $e$ is not
involved (as indicated by the absence of any symbol at the intersection of its
line with the vertical line representing the gate).  The action of gate 1
followed by gate 2 results in the following unitary transformation on the $xy$
basis vectors:
\begin{eqnarray}
  &&|xxx\rangle \to |xxx\rangle , \quad |xxy\rangle \to |yxy\rangle ,\quad
|xyx\rangle \to |xyx\rangle , \quad |xyy\rangle \to |yyy\rangle , \nonumber \\
&&|yxx\rangle \to |yyx\rangle , \quad |yxy\rangle \to |xyy\rangle ,\quad
|yyx\rangle \to |yxx\rangle , \quad |yyy\rangle \to |xxy\rangle .
\label{gates}
\end{eqnarray}

	If instead of the $xy$ basis, the $uv$ basis, see (\ref{euv}), is
employed for all three qubits, the same circuit, corresponding to the same
unitary transformation (\ref{gates}), takes the form shown in
Fig.~\ref{fgx}(b). The reason is that if {\it both} qubits involved in a
controlled-not gate are changed from the $xy$ to the $uv$ basis, the action of
the gate can again be represented as a controlled-not, but with the {\it
control and target qubits interchanged}, as the reader can easily verify using
(\ref{euv}).  Thus (\ref{gate1a}) is equivalent, again with qubit $a$ on the
left and $e$ on the right, to
\begin{equation}
  |uu\rangle \to |uu\rangle , \quad |uv\rangle \to |vv\rangle , \quad
|vu\rangle \to |vu\rangle , \quad |vv\rangle \to |uv\rangle ,
\label{gate1b}
\end{equation}
as can be checked by employing (\ref{euv}) with (\ref{gate1a}).  The result in
the $uv$ basis of the two gates acting in succession can be worked out either
by combining (\ref{euv}) with (\ref{gates}) or, more simply, by employing
(\ref{gate1b}) followed by the corresponding transformation for gate 2 in
Fig.~\ref{fgx}(b).

	The following is then an optimum strategy for Eve. She prepares qubits
$e$ and $f$ in initial states
\begin{eqnarray}
  |e_0\rangle = \sqrt{ 1-\Delta_{uv} }|x\rangle + \sqrt{\Delta_{uv}} |y\rangle
&=& \sqrt{ 1-D_{uv} }|u\rangle + \sqrt{D_{uv}} |v\rangle , \nonumber \\
|f_0\rangle = \sqrt{ 1-D_{xy} }|x\rangle + \sqrt{D_{xy}} |y\rangle &=&
\sqrt{1-\Delta_{xy} }|u\rangle + \sqrt{\Delta_{xy}} |v\rangle ,
\label{ef0}
\end{eqnarray}
where, with $w=uv$ or $xy$, $\Delta_w$ and $D_w$ are related through the
formulas:
\begin{equation}
 \Delta_w = \frac{1}{2} - \sqrt{D_w(1-D_w)},\quad D_w = \frac{1}{2} -
\sqrt{\Delta_w(1-\Delta_w)}.
\label{DDelta}
\end{equation}
We assume, for convenience, that both quantities are between 0 and 1/2, so that
(\ref{DDelta}) defines one quadrant of a circle of radius 1/2 centered at
$(1/2,1/2)$ in the $(D_w,\Delta_w)$ plane.  It is easy to check that the second
equality in each line in (\ref{ef0}) is consistent with (\ref{euv}) and
(\ref{DDelta}).

	After the initial preparation, qubits $e$ and $f$ interact with Alice's
qubit $a$ in the quantum circuit of Fig.~\ref{fgx}, and Eve allows $a$ to go on
to Bob while storing $e$ and $f$ until Alice announces the basis in which the
signal was transmitted.  Then Eve measures both qubits $e$ and $f$ in the basis
($xy$ or $uv$) announced by Alice.

	One can check that this is an optimum strategy by applying the unitary
transformation (\ref{gates}) to the initial state
\begin{equation}
 |a\rangle \otimes |e_0\rangle \otimes |f_0\rangle ,
\label{init}
\end{equation}
expressed as a linear combination of the $xy$ basis states, to obtain
$|X\rangle $ if $|a\rangle =|x\rangle $ and $|Y\rangle $ if $|a\rangle
=|y\rangle $, in the notation of the preceding paper, and $|U\rangle $ and
$|V\rangle $ by means of:
\begin{equation}
|U\rangle = (|X\rangle + |Y\rangle )/\sqrt{2}, \quad |V\rangle = (|X\rangle -
|Y\rangle )/\sqrt{2}.
\label{}
\end{equation}
Then using the projectors $\{E_\lambda\}$ and $\{F_\lambda\}$ defined in
Sec.~III of the preceding paper, one can verify that the conditions given there
(in Sec.~II) for saturating the bounds (\ref{Ixy}) and (\ref{Iuv}) are
satisfied.

	However, we shall use an alternative approach, which yields more
insight into the choice of coefficients in (\ref{ef0}): we shall calculate the
error rates and mutual information directly from the quantum circuit in
Fig.~\ref{fgx}, and verify that (\ref{Ixy}) and (\ref{Iuv}) are satisfied as
equalities.  Let us begin with the situation in which Alice announces, and Eve
measures in, the $xy$ basis, which can best be understood using
Fig.~\ref{fgx}(a).  First consider the case in which $D_{xy}$ and
$\Delta_{uv}$---note that Eve can choose them independently---are both equal to
zero, so that both $e$ and $f$ are initially in the state $|x\rangle $,
(\ref{ef0}).  Then gate 1 simply copies qubit $a$, $|x\rangle $ or $|y\rangle
$, to qubit $e$, so that by measuring $e$ in the $xy$ basis, Eve knows
precisely which signal Alice sent.  Furthermore, since $f$ is in the state
$|x\rangle $, qubit $a$ remains unchanged on its way from Alice to Bob, so
Eve's intervention causes no error.

	If $\Delta_{uv}=0$ but $D_{xy}$ is positive, Eve will again be able to
determine which signal Alice sent by measuring $e$.  However, measuring $f$
will yield $|y\rangle $ with probability $D_{xy}$ and $|x\rangle $ with
probability $1-D_{xy}$.  Since $f$ is the control qubit for gate 2, if it is in
state $|y\rangle $, an error will be produced in the transmission from Alice to
Bob, while if it is in state $|x\rangle $, there will be no error.  Hence Eve's
measurement of $f$, while it tells her nothing about which signal Alice sent,
shows her whether or not, in this particular case, Bob's measurement yielded
the same or the opposite result from what Alice transmitted.

	In the preceding paragraph we used a process of {\it retrodiction}, in
which we inferred the prior state of qubit $f$ from Eve's measurement.  This
can lead to quantum paradoxes when it is not used in the proper way, but in the
present context it can be justified, just as in \cite{gn96}, by using an
appropriate framework or family of consistent histories \cite{gr96}.  However,
rather than employing retrodiction for qubit $e$ as well, it is more
straightforward to adopt at the outset an appropriate consistent family, which
we shall call the $xy$ framework, based upon all three qubits being in either
an $|x\rangle $ or a $|y\rangle $ state at a time $t_1$ shortly after $t_0$,
when (\ref{ef0}) applies, but before qubits $a$ and $e$ reach the first gate,
see Fig.~\ref{fgx}, and at all later times as well~\cite{note1}. In this
framework, $e$ at time $t_1$ is in the state $|y\rangle $ with probability
$\Delta_{uv}$, and in $|x\rangle $ with probability $1-\Delta_{uv}$, the
absolute squares of the corresponding coefficients in (\ref{ef0}), while for
$f$ these probabilities are $D_{xy}$ and $1-D_{xy}$.  As long as we are using
the $xy$ framework, these probabilities can be thought of in the same way as in
a classical stochastic theory
\cite{gr96}.

	Because qubit $f$ at $t_1$, and thus at all later times until it is
measured, is in state $|y\rangle $ with probability $D_{xy}$, this is also the
probability of an error if Alice is transmitting to Bob in the $xy$ mode, as
noted earlier.  Next let us consider qubit $e$.  The results of Eve's
measurement, which is the value $e$ has when it leaves gate 1, will coincide
with the initial value of $a$ only if at $t_1$ $e$ is in the state $|x\rangle
$.  Otherwise the measured value will be the reverse of what Alice transmits.
Thus if one thinks of gate 1 as part of a communication channel from Alice
(qubit $a$) to Eve (qubit $e$), it is a noisy channel with a probability
$\Delta_{uv}$ that a bit will be flipped.  The mutual information $I_{xy}$
associated with such a channel is easily computed, assuming that Alice
transmits $|x\rangle $ or $|y\rangle $ with equal probability, and is given by
the right side of (\ref{Ixy}).

	To understand what happens when Alice sends a signal, and Eve makes her
measurements, in the $uv$ basis, we use Fig.~\ref{fgx}(b) and an alternative
consistent family, which we call the $uv$ framework, in which all three qubits
are in state $|u\rangle $ or $|v\rangle $ at $t_1$ and all later times
\cite{note2}.  Then in this framework, $e$ is in state $|v\rangle $ with
probability $D_{uv}$, and $|u\rangle $ with probability $1-D_{uv}$, at $t_1$;
see (\ref{ef0}). For $f$ the corresponding probabilities are $\Delta_{xy}$ and
$1-\Delta_{xy}$.  As $e$ is now the control qubit for gate 1,
Fig.~\ref{fgx}(b), it is at once obvious that when $e$ is in state $|v\rangle
$, there will be an error in the transmission from Alice to Bob when they use
the $uv$ basis.  Thus the error rate in this basis is $D_{uv}$, the probability
that $e$ is in state $|v\rangle $ at $t_1$, and hence at later times as well.
Also by measuring $e$ (in the $uv$ basis) Alice can determine whether or not
such an error has occurred.  However, measuring $e$ tells her nothing about
whether Alice sent a $|u\rangle $ or a $|v\rangle $.  To obtain this
information, she must measure qubit $f$.  The task is a bit more complicated
than in the case of the $xy$ basis considered earlier, because qubit $a$ may
have been flipped through its interaction with $e$ before it is copied to $f$.
However, since she also measures $e$, Eve can easily correct for this effect.
Consequently, the noise in the channel between Alice (qubit $a$) and Eve (qubit
$f$ corrected by $e$) is determined by the uncertainty in the value of $f$ at
$t_1$; a bit passing from Alice to Eve through this channel will be flipped
with probability $\Delta_{xy}$, the probability that $f$ is in state $|v\rangle
$ at $t_1$.  Again, the mutual information $I_{uv}$ for such a channel is
easily computed, and is given by the right side of (\ref{Iuv}).

	Consequently, one sees that the error rates produced by Eve's employing
the initial states in (\ref{ef0}) are, indeed, $D_{xy}$ and $D_{uv}$ in the
$xy$ and $uv$ bases, respectively, whereas the appropriate mutual information
in each case saturates the corresponding bound, (\ref{Ixy}) or (\ref{Iuv}).
This shows that Eve's strategy employing the gates in Fig.~\ref{fgx} is,
indeed, optimal.  Furthermore, one can understand how Eve faces a trade-off
between gaining information when Alice sends in one mode, and creating errors
when Alice uses the other mode. If Alice only employed the $xy$ mode, Eve
would, of course, set both $D_{xy}$ and $\Delta_{uv}$ equal to zero, as this
would cause no errors in the transmission from Alice to Bob, and produce a
perfect copy of Alice's signal in qubit $e$.  However, $\Delta_{uv}=0$ is
equivalent, (\ref{DDelta}), to $D_{uv}=1/2$, so that obtaining the maximum
possible information about the $xy$ transmission produces a large number of
errors in the $uv$ mode.  Similarly, setting $D_{xy}=0$, while it produces no
errors when Alice transmits in the $xy$ mode, has the consequence the
$\Delta_{uv}=1/2$, which means that Eve can extract no information whatever
when Alice transmits in the $uv$ mode.

	Eve's strategy as described above requires that she store {\it both}
qubits $e$ and $f$ while waiting for Alice to announce the basis used in
sending the signal.  Since the ``storage costs'' of preserving the qubits
against decoherence, should there be a long delay, could be high, it is worth
noting, as was pointed out in the preceding paper, that only one qubit needs to
be preserved if Eve just wants to estimate which signal Alice sent, and is not
interested in keeping track of whether an error occurred in the transmission
from Alice to Bob.  From Fig.~\ref{fgx}(a) it is evident that in the $xy$ case,
qubit $f$ could be discarded after it emerges from gate 2, since Eve only uses
$e$ to gain information about Alice's signal.  However, this would not work for
the $uv$ basis, where the information of interest is contained in the
correlation between the two qubits.  There are, nonetheless, two obvious
strategies available to Eve.  She can measure qubit $f$ in the $uv$ basis
immediately after it emerges from gate 2, and record the value in her notebook,
while preserving qubit $e$ for later analysis.  If Alice later announces that
she used the $xy$ basis, Eve ignores the record in her notebook, and measures
$e$ in the $xy$ basis.  If, on the other hand, the basis turns out to be $uv$,
Eve measures $e$ in that basis and uses it to correct the $f$ value she
measured earlier.  An alterative approach is to add a third gate to the circuit
in Fig.~\ref{fgx}, a controlled-not in which $e$ is the control and $f$ the
target in the $xy$ basis (and, of course, the reverse in the $uv$ basis).
After it passes through this third gate, Eve discards qubit $f$ and retains
qubit $e$ for later measurement in whichever basis is appropriate; this is
equivalent to the approach employed in the preceding paper.

	The use of the quantum circuit does not solve the problem of whether
Eve's optimum strategy is essentially unique.  One can show that interchanging
(a) and (b) in Fig.~\ref{fgx}, that is, employing the circuit in (a) for the
$uv$ basis and that in (b) for the $xy$ basis, is equivalent to the original
scheme preceded and followed by unitary transformations on the qubits of Eve's
probe, so that it is not different in any essential way.  However, this
obviously does not settle the question of uniqueness.

\section{Conclusion}
\label{s4}

	We have shown that Eve's optimum strategy can be represented by a
simple quantum circuit involving two controlled-not gates, along with the
preparation of the two qubits of her probe in suitable initial states, and
their later measurement in the same basis announced by Alice.  In this circuit,
the function of each qubit of the probe is clearly distinguished.  For example,
in the $xy$ basis, qubit $e$ is employed for extracting information about the
signal Alice sends, while $f$ creates errors in the transmission from Alice to
Bob.  While these errors can be reduced to zero by Eve's choice of a suitable
initial state for $f$, this choice makes it impossible for her to obtain any
information when Alice transmits in the $uv$ basis, for which Eve must extract
the information using $f$.

	While the use of the quantum circuit provides one with a certain amount
of insight into eavesdropping strategies, it does not by itself provide a proof
that the strategy is optimal, for which one needs the bounds derived in the
preceding paper, nor does it show that there is a unique optimal strategy.

\section*{Acknowledgements}
	We gratefully acknowledge helpful discussions with the other authors of
the preceding paper, and financial support for this research by NSF and ARPA
through grant CCR-9633102.

\begin{figure}[b]
\caption{Quantum circuit representing the interaction of Eve's probe,
qubits $e$ and $f$, with Alice's qubit $a$, in (a) the $xy$ basis and (b) the
$uv$ basis.  }
\label{fgx}
\end{figure}


\begin{thebibliography}{99}
\bibitem{note1}
	To be more precise, at $t_0$ we employ a decomposition of the identity
which includes projectors onto the states (\ref{init}) with $a$ set equal to
$x$ and $y$, and at all later times (except those at which two of the bits are
actually interacting inside one of the gate, and which are not included in the
histories) a decomposition of the identity onto the eight $xy$ basis states of
the three bits.
\bibitem{note2}
	The qualifications are the same as in \cite{note1}, except that $x$ and
$y$ are replaced by $u$ and $v$.  Also note that the the $uv$ framework is
incompatible with the $xy$ framework considered previously, so they cannot both
be applied simultaneously to the same system.  For further remarks on
incompatible frameworks, see \cite{gr96}.
\end{thebibliography}
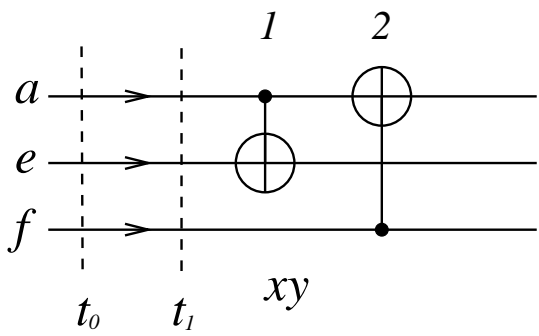
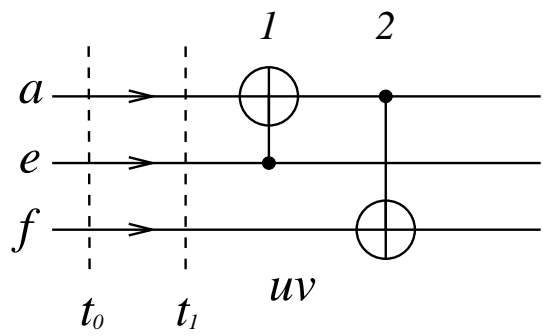
\end{document}